\documentclass{emulateapj}
\usepackage{amsmath}
\submitted{Accepted for publication in ApJ Letters on September 28, 2016}

\shorttitle{Damping of the Milky Way bar}

\shortauthors{E. L. {\L}okas}

\begin{document}

\title{Damping of the Milky Way bar by manifold-driven spirals}

\author{Ewa L. {\L}okas}
\affil{Nicolaus Copernicus Astronomical Center, Polish Academy of Sciences,
Bartycka 18, 00-716 Warsaw, Poland}

\begin{abstract}
We describe a new phenomenon of `bar damping' that may have played an important role in shaping the Milky Way bar and
bulge as well as its spiral structure. We use a collisionless $N$-body simulation of a Milky Way-like galaxy initially
composed of a dark matter halo and an exponential disk with Toomre parameter slightly above unity. In this
configuration, dominated by the disk in the center, a bar forms relatively quickly, after 1 Gyr of evolution. This is
immediately followed by the formation of two manifold-driven spiral arms and the outflow of stars that modifies the
potential in the vicinity of the bar, apparently shifting the position of the $L_1/L_2$ Lagrange points. This
modification leads to the shortening of the bar and the creation of a next generation of manifold-driven
spiral arms at a smaller radius. The process repeats itself a few times over the next 0.5 Gyr resulting in further
substantial weakening and shortening of the bar. The time when the damping comes to an end coincides with the first
buckling episode in the bar which rebuilds the orbital structure so that no more new spiral arms are formed. The
morphology of the bar and the spiral structure at this time show remarkable similarity to the present properties of the
Milky Way. Later on, the bar starts to grow rather steadily again, weakened only by subsequent buckling episodes
occurring at more distant parts of the disk.
\end{abstract}

\keywords{
Galaxy: formation --- galaxies: clusters: general --- galaxies: evolution --- galaxies: fundamental parameters
--- galaxies: kinematics and dynamics --- galaxies: structure}

\section{Introduction}

Bars are quite common morphological features among late-type galaxies. Depending on the exact definition about half
of them can be considered as barred \citep[see e.g.][and references therein]{Buta2015}. Bars formed in isolation are
believed to originate from instabilities in axisymmetric disks \citep[for a review see][]{Athanassoula2013} and their
evolution is controlled by a number of parameters, including the presence and properties of a live halo
\citep{Athanassoula2002, Debattista2006, Sellwood2016} and the initial velocity dispersion of the disk
\citep{Athanassoula2003}. In simulations, bars tend to grow in time in terms of length and strength while decreasing
their pattern speed. This rather steady growth can be slowed down by episodes of buckling instability that lead to the
vertical thickening of the bar and the formation of boxy/peanut shapes or pseudobulges \citep{Combes1990, Raha1991,
Martinez2006}. Bars can also form as a result of tidal interactions with other galaxies or environment \citep{Miwa1998,
Lang2014, Lokas2014, Lokas2016, Martinez2016}.

The Milky Way galaxy seems to be a typical barred spiral \citep[for a recent review of its
properties see][]{Bland2016}. The bar in the Milky Way is composed of a shorter (the length of 3 kpc) and a longer (5
kpc) component. The shorter one is thicker and known to possess a boxy/peanut shape since the observations in the
infrared by the COBE satellite \citep{Weiland1994}. The  density  distributions of the red clump stars within
1 kpc are approximately exponential with axis ratios 10:6.3:2.6 \citep{Wegg2013}. The bar pattern speed is not
accurately known and an average over different measurements gives values around $43 \pm 9$ km s$^{-1}$ kpc$^{-1}$. From
our position near the Sun we see the bar at an angle of about 27 deg. Recent kinematic measurements of the bar from the
APOGEE survey reveal that it is characterized by cylindrical rotation at the level of 170 km s$^{-1}$ at 35 deg of
galactic longitude and the central velocity dispersion of 120 km s$^{-1}$ \citep{Ness2016}.

The spiral structure of the Milky Way is much less constrained, mainly due to our unfavorable observational position
close to the disk plane. Although a number of spiral arms were identified and named, their detailed properties remain
poorly known. In a recent study \citet{Hou2014} combined data for thousands of different spiral tracers and found them
to be well fitted by models of both three- and four-arm logarithmic spirals. To further complicate the picture, they
also found that polynomial-logarithmic spirals (with variable pitch angles) are able to better match
the observed tangential directions. The origin of Milky Way spiral arms is equally vague and many scenarios have been
proposed, starting from the classical density wave theory of \citet{Lin1964} to spiral features seeded by density
inhomogeneities \citep{D'Onghia2013} and tidal effects of nearby dwarf galaxies \citep{Purcell2011}. It has also been
proposed that the spiral structure formation is driven by the bar through manifolds \citep{Athanassoula2009b}.

In this Letter we report on a discovery of a new phenomenon in the simulation of a bar in a Milky Way-like galaxy, which
we refer to as `bar damping'. In the configuration we consider here it occurs early on, soon after the formation of the
bar and seems to be tightly related to the manifold-driven spiral arms.

\section{The simulation}

The initial conditions for our simulation consisted of an $N$-body realization of a Milky
Way-like galaxy generated via procedures described in \citet{Widrow2005}
and \citet{Widrow2008} that allow to create $N$-body models of galaxies very near
equilibrium. Our Milky Way had parameters similar to the model MWa of \citet{Widrow2005}, with
two components: an NFW \citep{Navarro1997} dark matter halo and an exponential disk, but no classical bulge.
Each of the two components was made of $10^6$ particles.

The dark matter halo had a virial mass
$M_{\rm H} = 7.4 \times 10^{11}$ M$_{\odot}$ and concentration $c=18$. The disk had a mass $M_{\rm D}
= 5.5 \times 10^{10}$ M$_{\odot}$, about 25\% larger than the disk mass of MWa. We increased the disk mass by
the bulge mass of the original MWa model since we expect the (pseudo)bulge to form out of the disk as a
result of bar formation and evolution. The scale-length of the disk was $R_{\rm D} = 2.81$ kpc and
the thickness $z_{\rm D} = 0.41$ kpc. We assumed the central radial velocity dispersion of the disk
$\sigma_{R0}=121$ km s$^{-1}$. Both components of the galaxy were smoothly cut off at appropriate scales. The
initial disk mass fraction within $2.2 R_{\rm D}$ was 0.6 which means that our disk is close to maximal in agreement
with recent estimates for the Milky Way \citep{Bovy2013}. The minimum value of the Toomre parameter of this
realization was $Q=1.2$ so we expect the model to form a bar on a short time scale.

The evolution of the system in the simulation was followed for 10 Gyr using the GADGET-2 $N$-body code
\citep{Springel2001, Springel2005} with outputs saved every 0.05 Gyr. The adopted softening scales were $\epsilon_{\rm
D} = 0.1$ kpc and $\epsilon_{\rm H} = 0.7$ kpc for the disk and halo of the galaxy, respectively.

\begin{figure}
\begin{center}
    \leavevmode
    \epsfxsize=7.4cm
    \epsfbox[0 10 186 537]{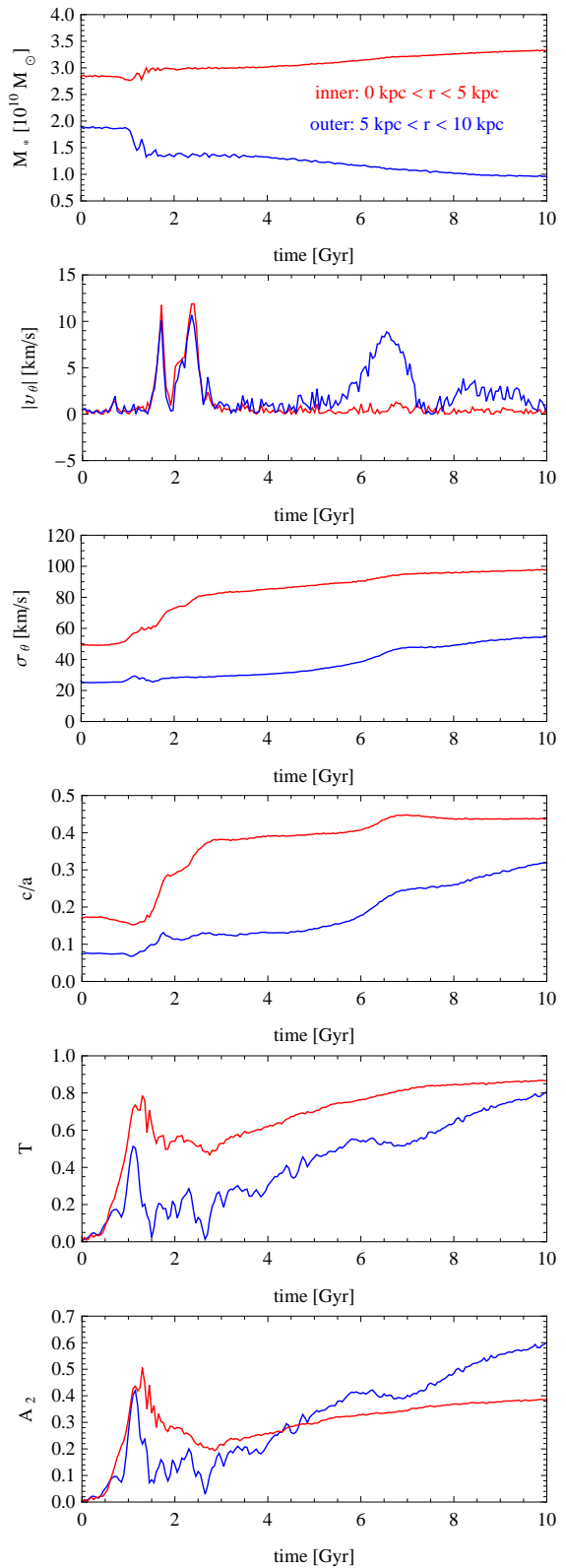}
\end{center}
\caption{Properties of the stellar component as a function of time from 0 to 10 Gyr. The measurements were done within
the radius of 5 kpc (red lines) and in the range 5 kpc $< r <10$ kpc (blue lines). The panels from top to bottom show
the stellar mass $M_{*}$, the absolute value of the mean velocity along the spherical coordinate $\theta$,
$|v_{\theta}|$, the velocity dispersion along $\theta$, $\sigma_{\theta}$, the ratio of the minor to major axis $c/a$,
the triaxiality parameter $T$ and the bar mode value $A_2$.}
\label{properties}
\end{figure}

\section{Evolution of the bar}

The evolution of the properties of the stellar component of the galaxy is illustrated in Figure~\ref{properties}. For
each simulation output we introduce the spherical coordinate system centered on the galaxy's center so that $\phi$
measures the angle in the disk plane and $\theta$ away from the plane. The measurements shown with the red lines were
done using stars within the spherical radius of 5 kpc and those shown with the blue lines using stars in the outer part
of the disk, within 5 kpc $ < r <10$ kpc. The upper panel gives the stellar mass contained in these radial ranges, the
next two panels show the kinematic properties in terms of the mean velocity along the spherical coordinate $\theta$,
$|v_{\theta}|$, and the velocity dispersion along $\theta$, $\sigma_{\theta}$. The last three panels plot the shape
properties of the stellar component in terms of the shortest to longest axis $c/a$, the triaxiality parameter
$T = [1-(b/a)^2]/[1-(c/a)^2]$ (where $a$, $b$ and $c$ are the lengths of the longest, intermediate and shortest axis of
the stellar component) and the bar mode $A_2$. The axis ratios were calculated using the inertia tensor and the $A_2$
from the projected positions of the stars onto the disk plane.

The evolution of the bar mode $A_2$ clearly shows that the bar starts to form very quickly after the beginning of the
simulation and its first maximum is reached soon after 1 Gyr from the start. After that the bar mode
abruptly decreases to start growing again around 3 Gyr from the beginning of the simulation. Similar conclusions are
supported by the analysis of the evolution of the triaxiality parameter: it increases strongly soon after 1 Gyr to
reach the values above $T=2/3$ in the inner part, signifying the formation of a strongly prolate spheroid. The moment
of the formation of the bar is accompanied by significant redistribution of stellar mass: as demonstrated by the first
panel, the stars are moved from the outer to the inner part of the galaxy. The axis ratio $c/a$ is rather stable over
the first Gyr but soon experiences dramatic increase. This increase is not smooth but happens in a number of rather
separate steps.

\begin{figure}
\begin{center}
    \leavevmode
    \epsfxsize=7.5cm
    \epsfbox[0 8 180 188]{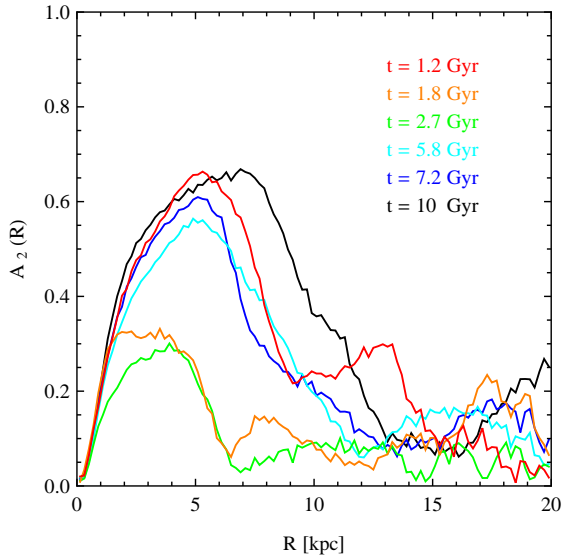}
\end{center}
\caption{The profiles of the bar mode $A_2 (R)$ at different times.}
\label{a2profiles}
\end{figure}

Since the thickening of the bar is usually caused by buckling, we have also calculated other, kinematic buckling
diagnostics in terms of the absolute value of the mean velocity along the spherical coordinate $\theta$,
$|v_{\theta}|$, and the velocity dispersion along $\theta$, $\sigma_{\theta}$. As discussed in \citet{Lokas2016}, the
presence of streaming velocity out of the disk plane (whether quantified by $|v_{\theta}|$ or $|v_z|$) is a good
signature of the presence of buckling instability. The effect of the buckling is to increase the $c/a$ axis ratio, but
also the velocity dispersion in the vertical direction, which we measure here by $\sigma_{\theta}$. Clearly, the
comparison between the third and fourth panels of Figure~\ref{properties} reveals that the two quantities trace each
other in both radial ranges. Interestingly, the behavior of $|v_{\theta}|$ is different in the inner and outer part of
the disk: while it has only two peaks in the inner part (red line), it shows four maxima in the outer part (blue line),
with every consecutive peak wider in time than the previous one. The analysis therefore points to only two buckling
episodes in the inner disk but four in the outer part.

Except for these multiple, rather than just one or two, buckling events typically seen in simulations
\citep{Martinez2006}, the evolution of the bar seems to proceed as usually found in similar studies. An
interesting behavior reveals itself however once we move from single-value measurements of characteristic properties of
the stellar component to more detailed ones. Since our main interest here is the bar properties the obvious
characteristic to calculate is the profile of the bar mode $A_2 (R)$. A few examples of such measurements along the
cylindrical radius $R$ at different times are shown in Figure~\ref{a2profiles}. The profile demonstrates
strong variability in time, from the highest values around 1.2 Gyr, when the maximum is $A_{2, {\rm max}}>0.6$ to much
smaller ones soon after, for example at $t= 1.8$ and 2.7 Gyr when $A_{2, {\rm max}}$ is only around 0.3. Later on the
bar mode grows again to reach the initial high values at the end of evolution.

An even more complete picture can be created if $A_2 (R)$ are calculated for each simulation output and plotted as a
function of time in the form of a map, such as the one shown in the upper panel of Figure~\ref{a2modestime}. This image
clearly reveals a kind of damping behavior of the bar mode between 1 and 1.7 Gyr from the start.

\section{Damping of the bar}

\begin{figure}
\begin{center}
    \leavevmode
    \epsfxsize=7.5cm
    \epsfbox[35 11.5 300 140]{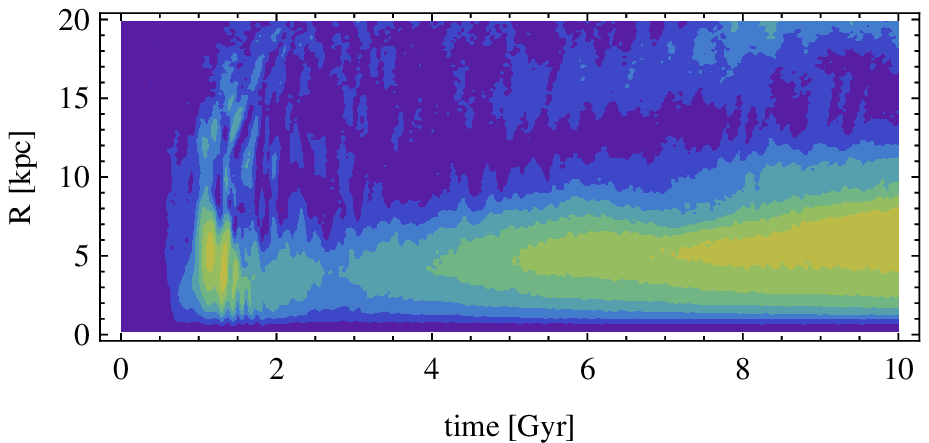}
\leavevmode
    \epsfxsize=0.837cm
    \epsfbox[38 -20 72 109]{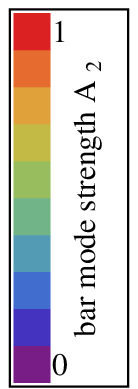}
\leavevmode
    \epsfxsize=7.5cm
    \epsfbox[35 11.5 300 140]{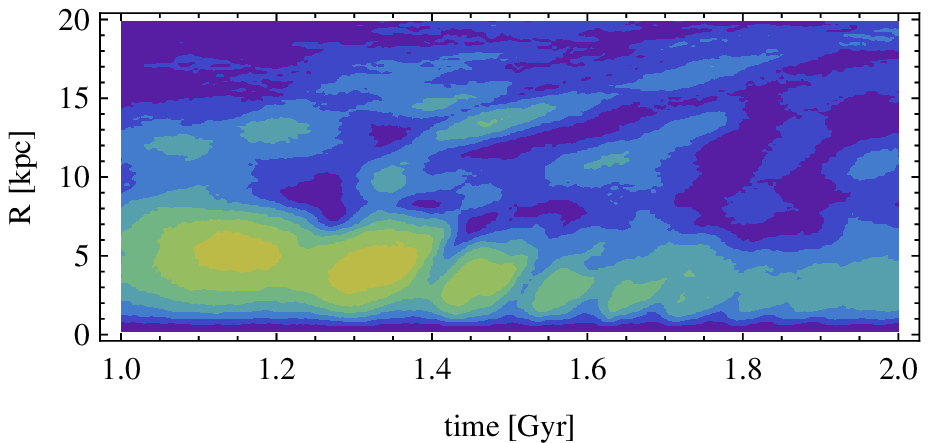}
\leavevmode
    \epsfxsize=0.837cm
    \epsfbox[38 -20 72 109]{legend2.eps}
\end{center}
\caption{The evolution of the profile of the bar mode $A_2(R)$ in time. The upper panel shows the evolution
over the time range of 0-10 Gyr, the lower one the evolution between 1 and 2 Gyr in more detail.}
\label{a2modestime}
\end{figure}

\begin{figure}
\begin{center}
\leavevmode
    \epsfxsize=4.2cm
    \epsfbox[0 0 250 270]{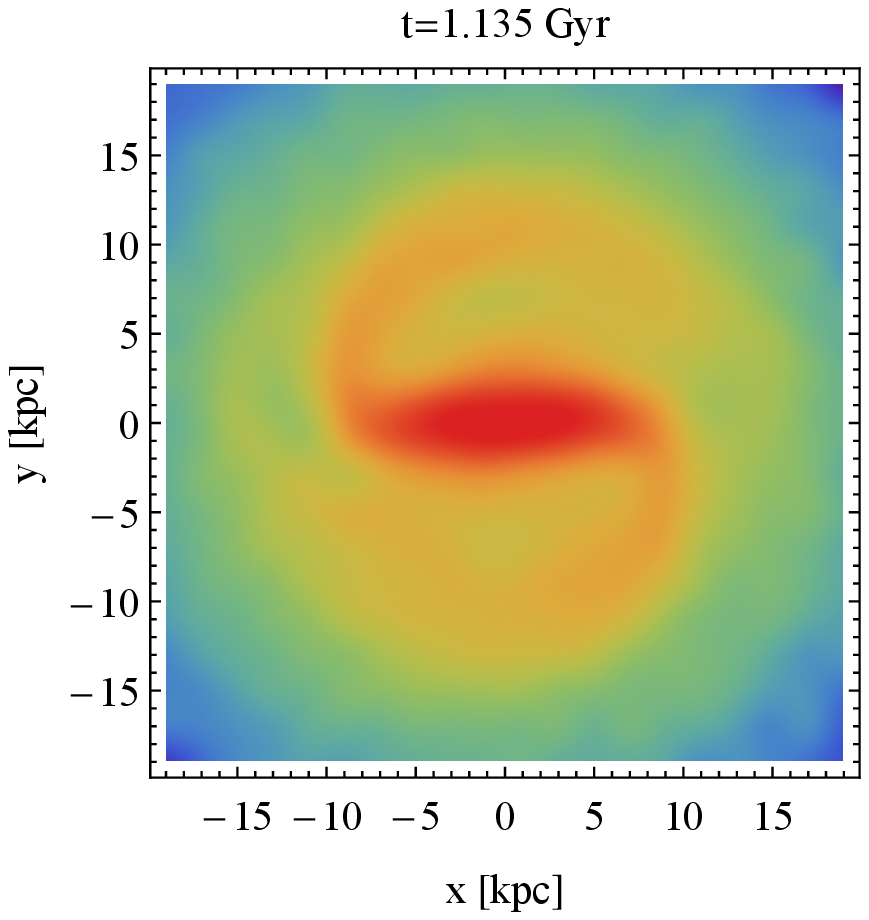}
\leavevmode
    \epsfxsize=4.2cm
    \epsfbox[0 0 250 270]{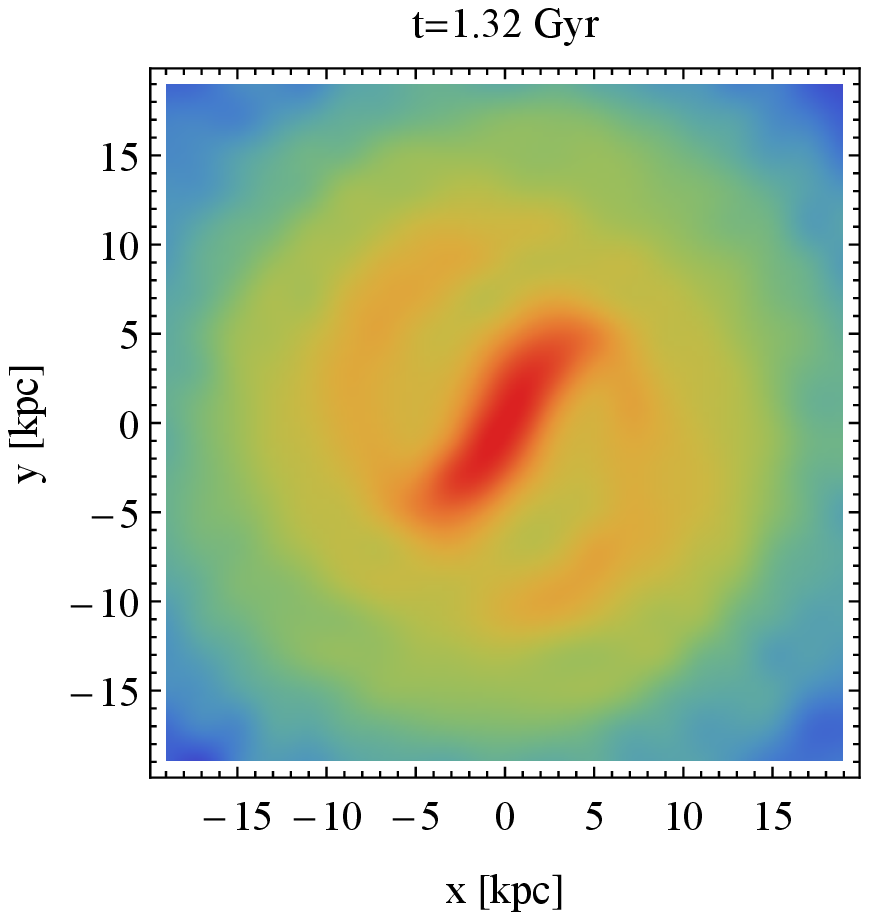}
\leavevmode
    \epsfxsize=4.2cm
    \epsfbox[0 0 250 270]{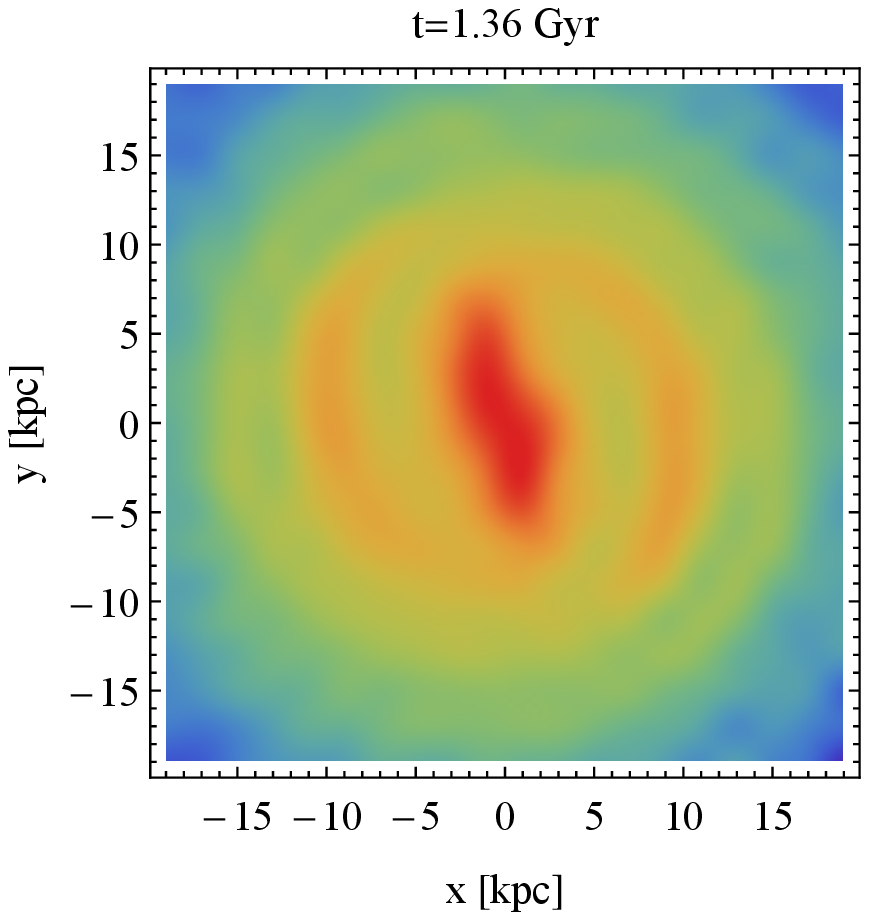}
\leavevmode
    \epsfxsize=4.2cm
    \epsfbox[0 0 250 270]{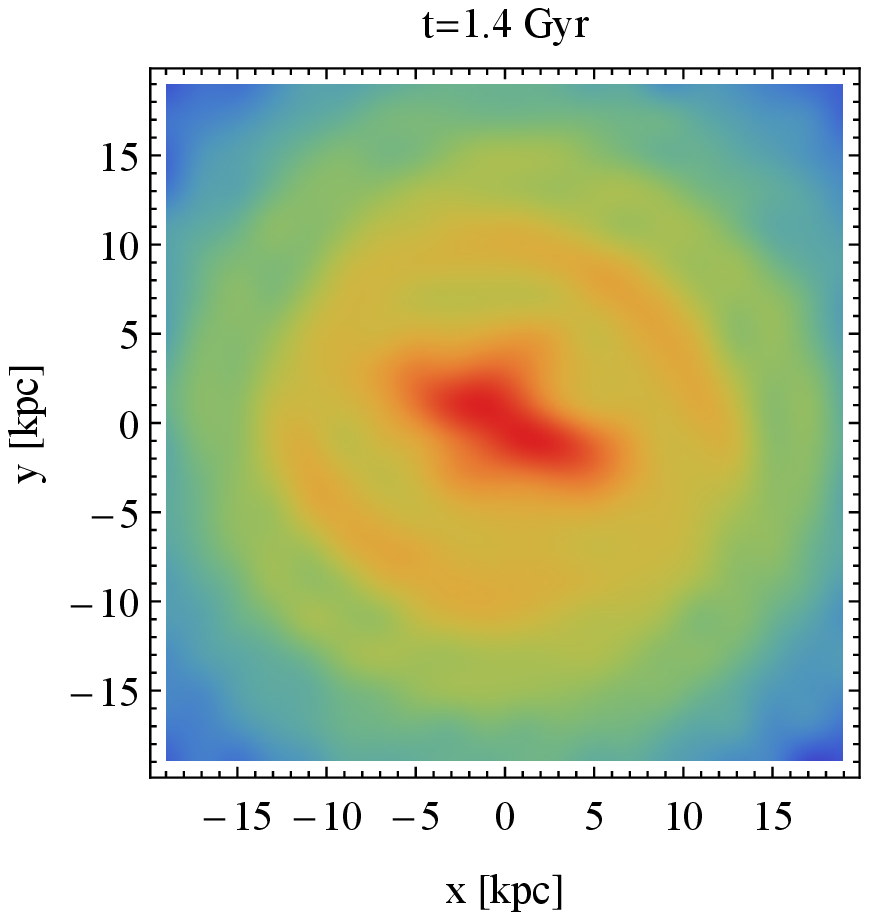}
\leavevmode
    \epsfxsize=4.2cm
    \epsfbox[0 0 250 270]{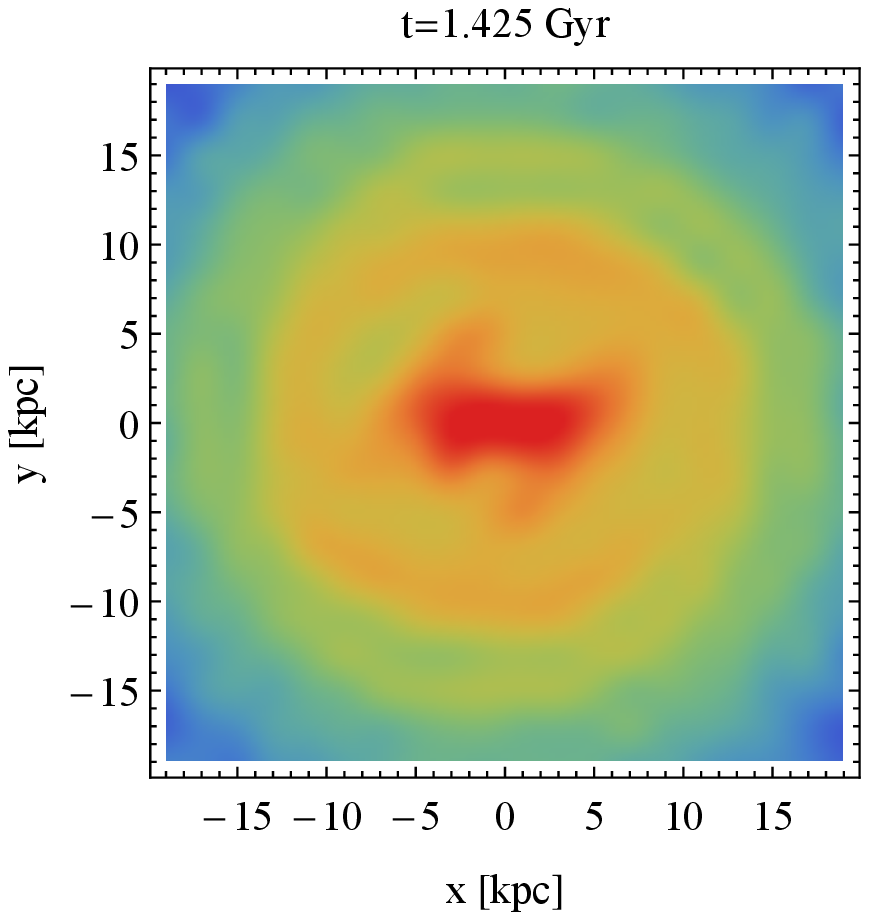}
\leavevmode
    \epsfxsize=4.2cm
    \epsfbox[0 0 250 270]{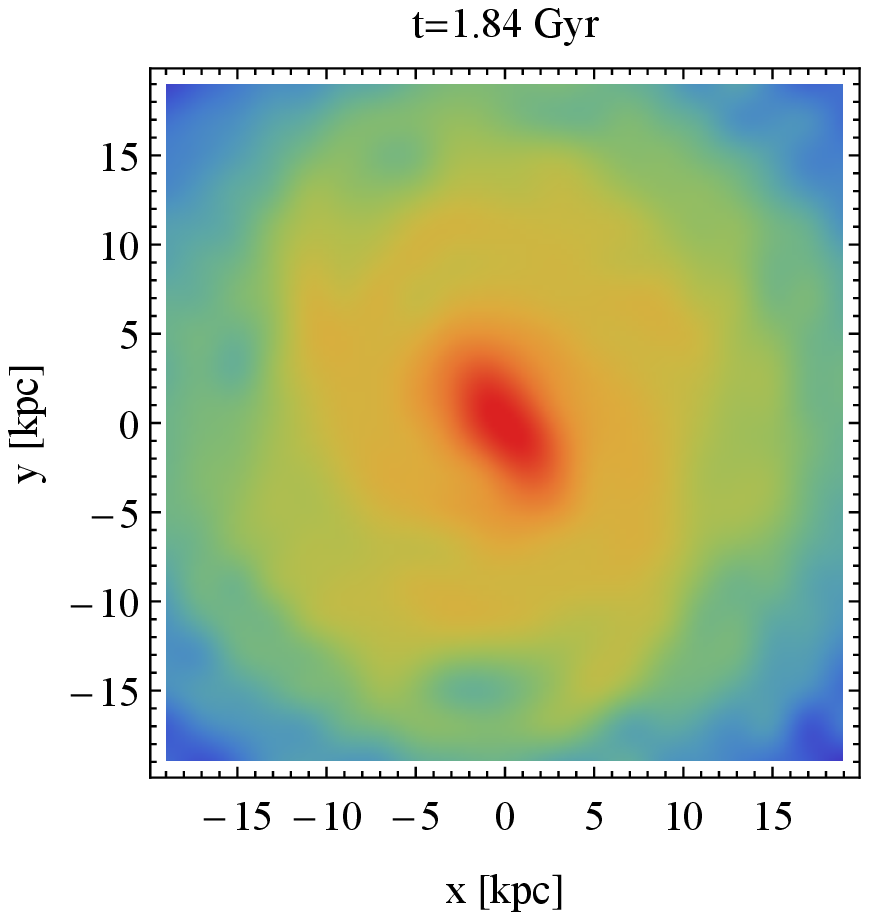}
\end{center}
\caption{Surface density distributions of the stars in the simulated galaxy viewed face-on at different times
marked at the top of each panel. The color scale was normalized to the maximum surface density at a
given time. The images illustrate different stages of the formation of manifold-driven
spiral arms at smaller and smaller radii. The lower right panel shows the galaxy image after the bar damping.}
\label{surden}
\end{figure}

In order to resolve this phenomenon better in time, we rerun the simulation between 1 and 2 Gyr saving outputs
ten times more often, i.e. every 0.005 Gyr. The map of the dependence of $A_2 (R)$ on time in
this time range is shown in the lower panel of Figure~\ref{a2modestime}. Now a clear pattern of repeated increase and
decrease of the bar mode is visible. To obtain some insight into the nature of the phenomenon we looked at the maps
of the surface density distribution of the stars viewed face-on. A few examples at different times are shown in
Figure~\ref{surden}. The plots show distinct images of the bar (not aligned with the $x$-axis but in the coordinate
system of the simulation box), but also reveal the presence of strong spiral arms or rings. The spiral arms clearly
seem to be driven by the bar and look like the manifold-driven spirals discussed by \citet{Athanassoula2012}.

\begin{figure}
\begin{center}
    \leavevmode
    \epsfxsize=7.8cm
    \epsfbox[0 10 186 515]{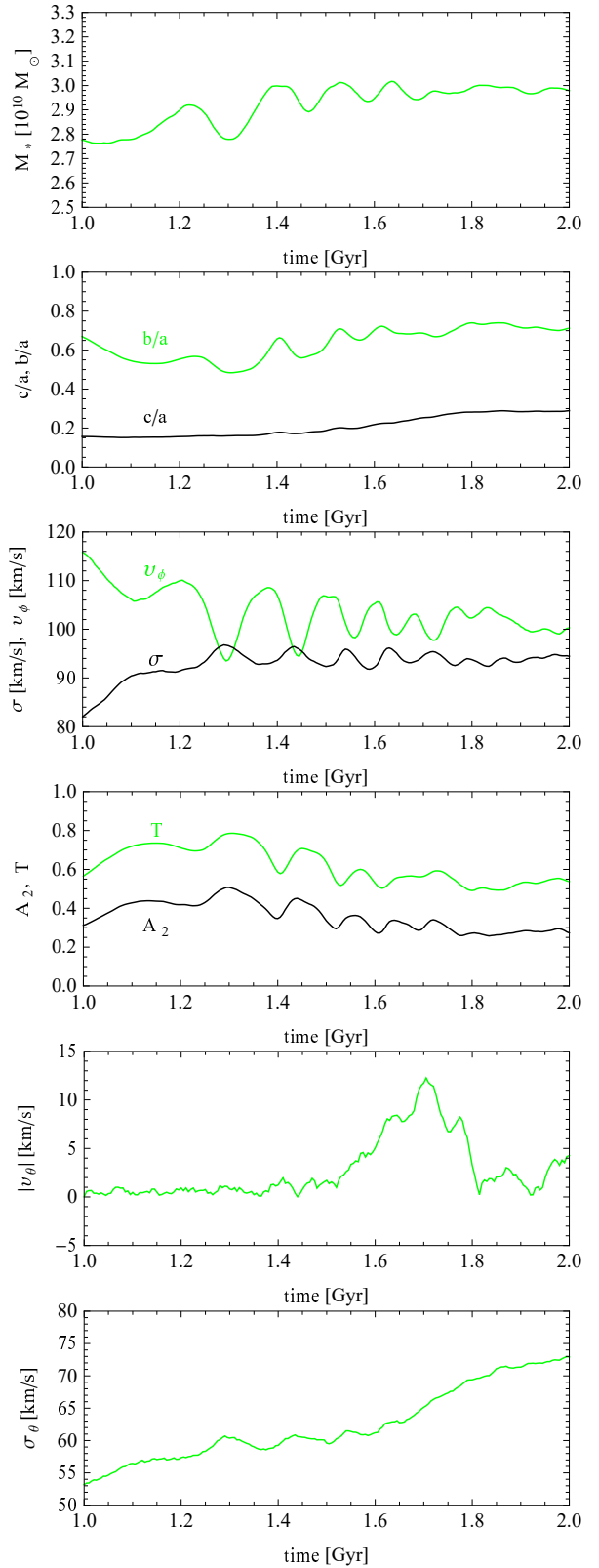}
\end{center}
\caption{Properties of the stellar component at the time of bar damping. The measurements were done using stars within
the radius of 5 kpc. The panels from top to bottom show the
stellar mass $M_{*}$, the axis ratios $c/a$ and $b/a$, the mean rotation velocity along the
spherical coordinate $\phi$, $v_{\phi}$, and the 1D velocity dispersion $\sigma$, the triaxiality parameter $T$
and the bar mode $A_2$, and the velocity and velocity dispersion along $\theta$, $|v_{\theta}|$ and $\sigma_{\theta}$.}
\label{propdetail}
\end{figure}

To study the phenomenon further we have measured the basic properties of the stellar component between 1 and 2 Gyr from
the start, with this higher time resolution. The measurements were done for stars within the radius of 5 kpc only this
time, because this is where the damping seems to occur and are shown in Figure~\ref{propdetail}. In addition to the
properties already discussed using Figure~\ref{properties}, namely the
stellar mass $M_{*}$ (first panel), the axis ratio $c/a$ (second panel), the triaxiality parameter $T$,
the bar mode $A_2$ (fourth panel), the velocity and velocity dispersion along $\theta$, $|v_{\theta}|$ and
$\sigma_{\theta}$ (the last two panels), we added here measurements of the intermediate to longest axis ratio $b/a$
(second panel), the mean rotation velocity along the spherical coordinate $\phi$, $v_{\phi}$ and the 1D velocity
dispersion $\sigma$ (third panel).

The oscillatory behavior of the quantities shown in the first four panels is clear. The outflow of stellar mass from
the inner region is accompanied by decreasing $b/a$, increasing $A_2$ and $T$, decreasing streaming velocity
$v_{\phi}$ while increasing the velocity dispersion $\sigma$, and vice versa, that corresponds to making the bar
stronger and then weaker. The amplitude of the variations decreases in time until about $t=1.6$ Gyr when the bar
experiences its first buckling episode as confirmed by the strong non-zero signal in $|v_{\theta}|$ and a significant
increase in $\sigma_{\theta}$ (two lower panels). The surface density maps shown in Figure~\ref{surden} covering the
time range between 1.135 and 1.84 Gyr confirm that the overall effect of the process is to significantly weaken and
shorten the bar.

\section{Discussion}

We interpret the observed behavior of the bar as due to the formation and evolution of manifold-driven spirals.
Manifolds were proposed as a possible origin of spiral arms and rings in barred galaxies \citep{Romero2006,
Romero2007, Athanassoula2009a, Athanassoula2009b, Athanassoula2010}. According to this theory, spiral arms are formed
by stars on orbits confined to manifolds associated with the periodic orbits around the saddle points of the potential
in the frame of reference corotating with the bar \citep{Binney2008}. In such a frame there are five equilibrium
(Lagrange) points at which the derivative of the effective potential vanishes. The most important for us here are the
$L_1$ and $L_2$ points located along the bar major axis, near its ends. These points are saddle points and unstable
causing the stars in their vicinity to escape from the neighborhood. Manifolds can be interpreted as tunnels along
which this escape can take place. The properties of manifolds so far were mostly studied by tracing orbits in
analytically set potentials, however, \citet{Athanassoula2012} demonstrated that they actually occur in full,
self-consistent $N$-body simulations and lead to the formation of spiral arms.

Here we have witnessed a similar phenomenon. However, in our case the formation of manifold-driven spirals is not a
single event but seems to repeat itself a few times. Around $t=1.135$ Gyr (see the upper left panel of
Figure~\ref{surden}) the manifold-driven spirals form for the first time. From the comparison between the
circular frequency of the galaxy and the pattern speed of the bar, $\Omega = \Omega_{\rm p}$, we can estimate the
corotation radius where the $L_1/L_2$ are located. For this time we obtain the value of $R_{\rm CR}=9$ kpc which
agrees with the distance from the center of the galaxy, where the spiral arms emanate from the bar. This confirms our
interpretation of the spirals as manifold-driven since according to the theory in cases of barred galaxies with no
rings and a relatively weak spiral structure the $L_1/L_2$ points are located at the end of the bar, where the arm
joins the bar \citep{Athanassoula2010}.

Later on the picture becomes more complicated. In addition to the outflow of stars along the outer manifolds
(spiral arms) there is a flow along the inner manifolds located near the bar \citep[see fig. 4 in][]{Athanassoula2010}
that rebuilds its structure so that for a short period of time it looks like a double bar (see the middle right panel
of Figure~\ref{surden}). The pattern speed of such a bar cannot be reliably measured so the position of the $L_1/L_2$
points is difficult to determine. However, these flows seem to modify the potential near the bar sufficiently in order
to move the $L_1/L_2$ points towards the center of the galaxy and thus decrease the length and strength of the bar.
This process could involve the stabilization of the $L_1/L_2$ points by the additional mass present in the
spiral arms and the creation of a new set of unstable $L_1/L_2$ points nearby as described in section 5 of
\citet{Athanassoula2009a}.

After the bar rebuilds itself it `emits' a next generation of spiral arms along outer manifolds
now located closer to the center and the bar is restructured. This shifts the position of the
$L_1/L_2$ points again and the whole process is repeated. One cycle of such evolution is shown by the four
panels corresponding to times $t=1.32$-1.425 Gyr in Figure~\ref{surden}. As illustrated in the lower panel of
Figure~\ref{a2modestime} and the first four panels of Figure~\ref{propdetail}, between 1 and 1.7 Gyr there are at least
five such cycles distinguishable. At about $t=1.6$ Gyr the bar starts to buckle for the first time so its orbital
structure is substantially rebuilt and no well-defined bar-driven spiral arms form later on.

Interestingly, right after the damping period is finished the bar is quite short, with the length of the order of 5
kpc, as can be estimated from the drop of the $A_2(R)$ profile to half its maximum value (see the orange line for
$t=1.8$ Gyr in Figure~\ref{a2profiles}). In addition, it is surrounded by a family of rather irregular spiral arms of
different length originating from the multiple generations of manifold-driven double spirals (see the lower right panel
of Figure~\ref{surden}). This morphology is qualitatively similar to the present spiral structure of the Milky Way as
far as we know it \citep{Vallee2008, Hou2014}. We note that this transition from two- to multiple-arm structure
cannot be explained by arguments based on the swing amplification theory and the disk stability criteria
\citep{Athanassoula1987, DOnghia2015}. According to this theory the number of arms grows with decreasing disk mass
fraction within $2.2 R_{\rm D}$ while in our case this fraction remains approximately constant (within 2\%) during
the whole damping period and close to the initial value of 0.6.

We conclude that the mechanism of bar damping we described here may have
contributed to the formation of the present-day structure of the Milky Way. It could influence the shaping
of the bar and spiral structure, as these seem to be intimately related in this process, but may also have an effect on
the formation of Milky Way (pseudo)bulge if it formed via buckling instability as the first episode of this
phenomenon takes place immediately after damping. Although in our simulation the damping occurs early in the evolution
of the bar, it is quite possible that with different initial conditions and/or addition of gas physics it could be a
much more recent phenomenon.

\section*{Acknowledgments}

This work was supported in part by the Polish National Science Centre under grant 2013/10/A/ST9/00023.
We thank L. Widrow for providing procedures to generate $N$-body realizations for initial conditions.
Useful comments from an anonymous referee are kindly appreciated.

\end{document}